%% file: d0-bs-lifetime.tex
\documentclass[aps,prl,twocolumn,showpacs,superscriptaddress,10pt,floatfix]{revtex4}
\usepackage{graphicx}
\usepackage{dcolumn}
\usepackage{bm} 
\usepackage{amssymb}
\usepackage{amsmath}
\usepackage{epsfig}
\usepackage{subfigure,tabularx}

\begin{document}

\hspace{5.2in} \mbox{FERMILAB-PUB-14-388-E}

\title{Measurement of the \boldmath{$B_s^0$} Lifetime in the Flavor-Specific Decay Channel \boldmath{$B_s^0 \to D_s^- \mu^+\nu X$}}

\input author_list.tex

\date{Received 7 October 2014; revised manuscript received 1 December 2014; published 9 February 2015}
\begin{abstract}

We present an updated measurement of the $B_s^0$ lifetime using the semileptonic decays 
$B_s^0\rightarrow D_s^-\mu^+\nu X$, with $D_s^- \to \phi \pi^-$ and $\phi \to K^+K^-$ (and the charge conjugate 
process). This measurement uses the full Tevatron Run II sample of proton-antiproton collisions at 
$\sqrt{s} = 1.96$~TeV, comprising an integrated luminosity of 10.4~fb$^{-1}$. 
We find a flavor-specific lifetime 
$\tau_{\mathrm{fs}}(B_s^0)=1.479\pm0.010\thinspace{\rm(stat)}\pm0.021\thinspace{\rm(syst)}\thinspace\rm{ps}$. 
This technique is also used to determine the $B^0$ lifetime using the analogous 
$B^0\to D^-\mu^+\nu X$ decay with $D^-\to\phi\pi^-$ and $\phi\to K^+K^-$, yielding  
$\tau(B^0)=1.534\pm0.019\thinspace{\rm(stat)}\pm0.021\thinspace{\rm(syst)}\thinspace\rm{ps}$. 
Both measurements are consistent with the current world averages, and the $B_s^0$ lifetime measurement is 
one of the most precise to date. Taking advantage of the cancellation of systematic uncertainties, we determine the 
lifetime ratio $\tau_{\mathrm{fs}}(B_s^0)/\tau(B^0) = 0.964\pm0.013\thinspace{\rm(stat)}\pm0.007\thinspace{\rm(syst)}$.

\end{abstract}

\maketitle

The decays of hadrons containing a $b$ quark are dominated by the weak interaction of the $b$ quark.
In first-order calculations, the decay widths of these hadrons are independent of the 
flavor of the accompanying light quark(s).
Higher-order predictions break this symmetry, with the spectator quarks having roles in the time evolution 
of the $B$ hadron decay~\cite{theory1,theory2}.
 The flavor dependence leads to an expected lifetime hierarchy of
$\tau(B_c)<\tau(\Lambda_b)<\tau(B_s^0) \approx \tau(B^0) < \tau(B^+)$, which has been observed experimentally~\cite{pdg}.
The ratios of the lifetimes of different $b$ hadrons
are precisely predicted by heavy quark effective theories 
and provide a way to 
experimentally study these higher-order effects, and to test for possible new physics beyond the standard model~\cite{bobeth}.
Existing measurements are in excellent agreement with predictions~\cite{pdg} for the lifetime ratio $\tau(B^+)/\tau(B^0)$, 
but until recently the experimental precision has been insufficient to test the corresponding theoretical prediction for
$\tau(B_s^0)/\tau(B^0)$. 
In particular, predictions using inputs from unquenched lattice QCD calculations give
$0.996 < \tau(B_s^0)/\tau(B^0) < 1$~\cite{theory2}.
More precise measurements of both $B_s^0$ lifetime and the ratio to its lighter counterparts 
are needed to test and refine the models. 

A flavor-specific final state such as $B_s^0\to D_s^-\mu^+\nu$ is one where the charges of the 
decay products can be used to know whether the meson was a $B^0_s$ or $\bar{B^0_s}$ at the time of decay.
As a consequence of neutral $B$ meson flavor oscillations, the $B_s^0$ lifetime as measured 
in semileptonic decays 
is actually a combination of the lifetimes of the heavy and light mass eigenstates with an equal mixture of these 
two states at time $t=0$.  If the resulting superposition of two exponential distributions is fitted with a 
single exponential function, one obtains to second order~\cite{theory4}
\begin{eqnarray}
\tau_{\mathrm{fs}}(B_s^0) = \frac{1}{\Gamma_s} \cdot \frac{1 + (\Delta \Gamma_s / 2\Gamma_s)^2}{1 - (\Delta \Gamma_s / 2\Gamma_s)^2},
\end{eqnarray}
where $\Gamma_s = (\Gamma_{sL} + \Gamma_{sH})/2$ is the average decay width of the light and heavy states, 
and $\Delta\Gamma_s$ is the difference $\Gamma_{sL} - \Gamma_{sH}$.
This dependence makes the flavor-specific lifetime an important parameter in global fits~\cite{hfag} used to extract 
$\Delta\Gamma_s$, and hence, to constrain possible $CP$ violation in the mixing and interference of $B_s^0$ mesons.

Previous measurements have been performed by the CDF~\cite{cdf}, D0~\cite{slprl}, and 
LHCb~\cite{lhcb,lhcb2} Collaborations, with additional earlier measurements from LEP~\cite{lep}.
During Run II of the Tevatron collider from $2002$--$2011$, the D0 detector~\cite{d0det} accumulated 10.4~fb$^{-1}$ of
$p\bar{p}$ collisions at a center-of-mass energy of $1.96~\mathrm{TeV}$.
We present a precise measurement of the $B_s^0$ lifetime that uses the flavor-specific decay 
$B_s^0\rightarrow D_s^-\mu^+\nu X$, with $D_s^- \to \phi \pi^-$ and $\phi \to K^+K^-$~\cite{conjugate},
selected from this dataset. It is superseding our previous measurement~\cite{slprl}.

A detailed description of the D0 detector can be found elsewhere~\cite{d0det}.
The data for this analysis were collected with a single muon trigger. 
Events are considered for selection if they contain a muon candidate identified through signatures 
both inside and outside the toroid magnet~\cite{d0det}. 
The muon must be associated with a central track, have transverse momentum ($p_T$) exceeding 2.0~GeV$/c$, 
and a total momentum of $p > 3.0$~GeV$/c$.
Candidate $B_s^0 \to D_s^-\mu^+\nu X$ decays
are reconstructed by first combining two charged particle tracks of opposite charge,
which are assigned the charged kaon mass. 
Both tracks must satisfy $p_T > 1.0~\mathrm{GeV/}c$, and the invariant mass of the two-kaon system 
must be consistent with a $\phi$ meson, $1.008~\mathrm{GeV/}c^2 < M(K^+K^-) < 1.032~\mathrm{GeV/}c^2$.
This $\phi$ candidate is then combined with a third track, assigned the charged pion mass, 
to form a $D_s^- \to \phi \pi^-$ candidate.
The pion candidate must have $p_T > 0.7~\mathrm{GeV/}c$, and the invariant mass of the $\phi \pi^-$ system 
must lie within a window that includes the $D_s^-$ meson, $1.73~\mathrm{GeV/}c^2 < M(\phi\pi^-) < 2.18~\mathrm{GeV/}c^2$.
The combinatorial background is reduced by requiring that the three tracks create a common $D_s^-$ vertex as described
in Ref.~\cite{durham}.
Lastly, each $D_s^-$ meson candidate is combined with the muon to reconstruct a $B_s^0$ candidate.
The invariant mass must be within the range $3~\mathrm{GeV/}c^2 < M(D_s^-\mu^+) < 5~\mathrm{GeV/}c^2$.
All four tracks must be associated with the same $p\bar{p}$ interaction vertex (PV), 
and have hits in the silicon and fiber tracker detectors.

Muon and pion tracks from genuine $B_s^0$ decays must have opposite charges, 
which defines the right-sign sample.
The wrong-sign sample is also retained to help constrain the background model.
In the right-sign sample, the reconstructed $D_s^-$ meson is required to be displaced from 
the PV in the same direction as its momentum in order to reduce background.

The flavor-specific $B_s^0$ lifetime, $\tau({B_s^0})$, can be related to the decay kinematics in the transverse plane,
$c\tau({B_s^0}) = L_{xy}M/p_T(B_s^0)$, 
where $M$ is the $B_s^0$ mass, taken as the world average~\cite{pdg},
and $L_{xy} = \vec{X}\cdot\vec{p}_T/|\vec{p}_T|$ is the transverse decay length, 
where $\vec{X}$ is the displacement vector from the PV
to the secondary vertex in the transverse plane.
Since the neutrino is not detected, and the soft hadrons and photons from decays of 
excited charmed states are not explicitly included in the reconstruction, the $p_T$ of the
$B_s^0$ meson cannot be fully reconstructed. 
Instead, we use the combined $p_T$ of the muon and $D_s^-$ meson, $p_T(D_s^-\mu^+)$. 
The reconstructed parameter is the pseudoproper decay length, 
PPDL~$= L_{xy}M/p_T(D_s^-\mu^+)$.
To model the effects of the missing $p_T$ and of the momentum resolution when the $B_s^0$ lifetime is extracted from the PPDL distribution, 
a correction factor $K$ is introduced, defined by $K = p_T(D_s^-\mu^+)/p_T(B_s^0)$.
It is extracted from a Monte Carlo (MC) simulation, separately for 
a number of specific decays comprising both signal and background components.

MC samples are produced using the {\sc pythia} event generator~\cite{pythia} to model the production 
and hadronization phase, interfaced with {\sc evtgen}~\cite{evtgen} to model the decays of $b$ and $c$ hadrons. 
The events are passed through a detailed {\sc geant} simulation of the detector~\cite{geant}
and additional algorithms to reproduce the effects of digitization, detector noise, and pileup. 
All selection cuts described above are applied to the simulated events.
To ensure that the simulation fully describes the data, and in particular to account for the effect of muon triggers,
we weight the MC events to reproduce the muon transverse momentum distribution observed in data. 

\begin{table}[htb]
\begin{tabularx}{\columnwidth}{l>{\raggedleft\arraybackslash}X}
\hline\hline
Decay channel & Contribution \\
\hline
$D_s^-\mu^+\nu_{\mu}$                                                                    & $(27.5\pm2.4)\%$ \\
$D_s^{*-}\mu^+\nu_{\mu}\times(D_s^{*-}\rightarrow D_s^-\gamma/D_s^-\pi^0)$               & $(66.2\pm4.4)\%$ \\
$D_{s(J)}^{*-}\mu^+\nu_{\mu}\times(D_{s(J)}^{*-}\rightarrow D_s^{*-}\pi^0/D_s^{*-}\gamma)$  \hfill & $(0.4\pm5.3)\%$  \\
$D_s^{(*)-}\tau^+\nu_{\tau}\times (\tau^+\rightarrow \mu^+\bar{\nu}_{\mu}\nu_{\tau})$    & $(5.9\pm2.7)\%$  \\
\hline\hline
\end{tabularx}
\caption{Relative contributions to the $D_s^-\mu^+$ signal from different semileptonic $B_s^0$ decays.
The uncertainties are dominated by limited knowledge of the branching fractions~\cite{pdg,evtgen}.
In total, these processes comprise $(80.5\pm2.1)\%$ of the events in the $D_s^-\mu^+$ mass broad peak after 
subtracting combinatorial background.}
\label{table1}
\end{table}

Table~\ref{table1} summarizes the semileptonic $B_s^0$ decays that contribute to the $D_s^-\mu^+$ signal.
Experimentally these processes differ only in the varying amount 
of energy lost to missing decay products, which is reflected in the final $K$-factor distribution.
Table~\ref{table2} shows the 
list of non-negligible processes from subsequent semileptonic charm decays which also contribute 
to the signal.
These two tables represent the sample composition of the $D_s^-\mu^+$ signal.

\begin{table}[htb]
\begin{tabularx}{\columnwidth}{l>{\raggedleft\arraybackslash}X}
\hline\hline
Decay channel              & Contribution \\
\hline
$B^+\rightarrow D_s^-DX$              & $(3.81\pm0.75)\%$ \\
$B^0\rightarrow D_s^-DX$              & $(4.13\pm0.70)\%$ \\
$B_s^0\rightarrow D_s^-D_s^{(*)}X$    & $(1.11\pm0.36)\%$ \\
$B_s^0\rightarrow D_s^-DX$            & $(0.92\pm0.44)\%$ \\
$c\bar{c}\rightarrow D_s^-\mu^+$      & $(9.53\pm1.65)\%$ \\
\hline\hline
\end{tabularx}
\caption{Other semileptonic decays contributing to the $D_s^-\mu^+$ signal.
Listed contributions are obtained after
subtracting combinatorial background. The uncertainties are dominated by limited knowledge of the
branching fractions~\cite{pdg,evtgen}.}
\label{table2}
\end{table}

\begin{figure}[htb]
  \centering
  \includegraphics[width=\columnwidth]{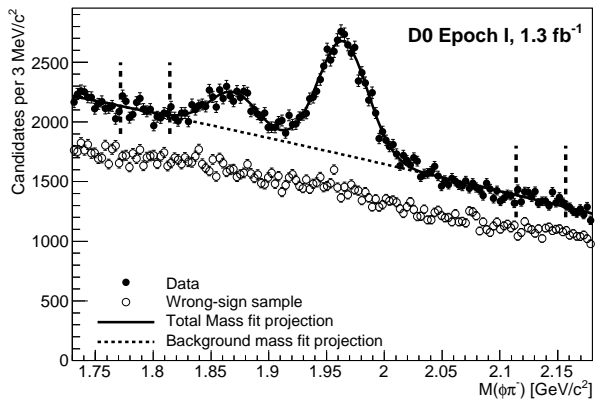}
  \caption[]{Distributions of the invariant mass $M(\phi\pi^-)$ for $D_s^-\mu^+$ candidates passing all selection
criteria in one of the five data periods. The higher-mass peak is the $D_s^-$ signal, with a smaller $D^-$ peak at lower mass.
Sidebands for right-sign sample are indicated with dashed lines  and the corresponding 
distribution for the wrong-sign sample is also shown.}
\label{fig1}
\end{figure}

We partition the dataset into five data-collection periods,
separated by accelerator shutdowns, each comprising $1$--$3$~fb$^{-1}$ of integrated luminosity,
to take into account time- or luminosity-dependent effects.
The behavior and overall contribution of the dominant combinatorial backgrounds changed 
as the collider, detector, and trigger conditions evolved over the course of the Tevatron Run II.
Figure~\ref{fig1} shows the $M(\phi \pi^-)$ invariant mass distribution for the right-sign $D_s^- \mu^+$ candidates
for one of these data periods.
Lifetimes are extracted separately for each period; they are
consistent within uncertainties and a weighted average is made for the final measurement.
The MC weighting as a function of $p_T$ is performed
separately for each of the five data samples. The $K$ factors are extracted independently in each sample, 
with significant shifts observed due to the changing trigger conditions. The $K$-factor distribution peaks 
at $\approx 0.9$ for the $D_s^-$ signal and at $\approx 0.8$ for the first four backgrounds listed in Table~\ref{table2}.  
The $K$-factor  distribution populates $0.5 < K < 1$ for both the signal and background components.

To determine the number of events in the signal region and define the signal and background samples, 
we fit a model to the $M(\phi\pi^-)$ invariant mass distribution as shown in Fig.~\ref{fig1}.
The $D_s^-$ and $D^-$ mass peaks are each modeled using an independent Gaussian distribution 
to represent the detector mass resolution, and a second-order polynomial is used to model the combinatorial background. 
Using the information obtained from these fits, we define the signal sample (SS) as those events in the 
$M(\phi\pi^-)$ mass distribution that are within $\pm 2\sigma$ of the fitted mean $D_s^-$ meson mass, 
where $\sigma$ is the Gaussian width of the $D_s^-$ mass peak obtained from the fit. 
We find  a total of $72\thinspace 028 \pm 727$ $D_s^-\mu^+$ signal events in the full dataset.
Yields observed in the different periods are consistent with expectations
taking into account changing trigger conditions and detector performance.
The background sample (BS) includes
those events in the sidebands of the $D_s^-$ mass distribution given by
$-9\sigma$ to $-7\sigma$ and $+7\sigma$ to $+9\sigma$ from the fitted  mean mass. 
Wrong-sign events in the full $M(\phi\pi^-)$ range are also included in the background sample, yielding more events
to constrain the behavior of the combinatorial background. 

The extraction of the flavor-specific $B_s^0$ lifetime is performed using an unbinned maximum likelihood fit 
to the data, based on the PPDL of each candidate~\cite{jorgethesis}. 
The effects of finite $L_{xy}$ resolution of the detector and the $K$ factors 
are included in this fit to relate the underlying decay time of the candidates 
to the corresponding observed quantity.
The signal and background samples defined above are fitted simultaneously, with a single 
shared set of parameters used to model the combinatorial background shape.
To validate the lifetime measurement method, we perform a simultaneous fit of the 
$B^0$ lifetime using the Cabibbo suppressed  decay $B^0\rightarrow D^-\mu^{+} X$ seen
in Fig.~\ref{fig1} at lower masses. 
This measurement also enables the ratio $\tau_{\rm{fs}}({B^0_s})/\tau({B^0})$ to be measured with high precision,
since the dominant systematic uncertainties are highly correlated between the two lifetime measurements. 
For simplicity, the details of the fitting function are illustrated for the $B_s^0$ lifetime fit alone. 
In practice an additional likelihood product is included to extract the $B^0$ lifetime in an identical manner.

The likelihood function $\mathcal{L}$ is defined as
\begin{equation}
\mathcal{L} = \prod_{i\in \text{SS}}[ f_{D_s\mu}\mathcal{F}_{D_s\mu}^{i} + 
(1-f_{D_s\mu})\mathcal{F}_{\text{comb}}^{i}] \prod_{j\in \text{BS}}\mathcal{F}_{\text{comb}}^{j}\thinspace , 
\label{pdf}
\end{equation}
\noindent where 
$f_{D_s\mu}$ is the fraction of $D_s^-\mu^+$ candidate events in the signal sample, obtained from the fit 
of the $D_s^-$ mass distribution, and $\mathcal{F}_{D_s\mu\text{(comb)}}^{i}$ is the candidate 
(combinatorial background) probability density function (PDF) evaluated for the $i{\text{th}}$ event. 
The probability density $\mathcal{F}_{D_s\mu}^i$ is given by
\begin{eqnarray}
\mathcal{F}_{D_s\mu}^i  
  &=& f_{\bar{c}c}F_{\bar{c}c}^i + f_{B1}F_{B1}^i + f_{B2}F_{B2}^i + f_{B3}F_{B3}^i  + f_{B4}F_{B4}^i \nonumber \\
  &+& \Big(1-f_{\bar{c}c} - f_{B1} - f_{B2} - f_{B3} - f_{B4}\Big)F_s^i. 
\label{SigLL}
\end{eqnarray}
\noindent Each factor $f_X$ is the expected fraction of a particular component $X$ in the signal sample, 
obtained from simulations and listed in Tables~\ref{table1} and \ref{table2}.
The first term accounts for the prompt $c\bar{c}$ component, and the decays 
$B1$--$B4$ represent the first four components listed in Table~\ref{table2}.
The last term of the sum in Eq.~(\ref{SigLL}) represents the signal events 
$S \equiv (B_s^0\rightarrow D_s^-\mu^+\nu X)$ listed in Table \ref{table1}.
The factor $F_{\bar{c}c}$ is the lifetime PDF for the $\bar{c}c$ events, given by a Gaussian distribution 
with a mean of zero and a free width. 
Each $B$ decay mode is associated with a separate PDF, $F_{X}$, modeling the PPDL distribution, 
given by an exponential decay convoluted with a resolution function and 
with the $K$-factor distribution. 
All $B$-meson decays
are subject to the same PPDL resolution function.
A double-Gaussian distribution is used for the resolution function, with widths given 
by the event-by-event PPDL uncertainty determined from the 
$B^0_s$ candidate vertex fit multiplied by two overall scale factors and a ratio
between their contributions that are all allowed to vary in the fit.

\begin{figure}[htb]
  \centering           
\includegraphics[width=\columnwidth]{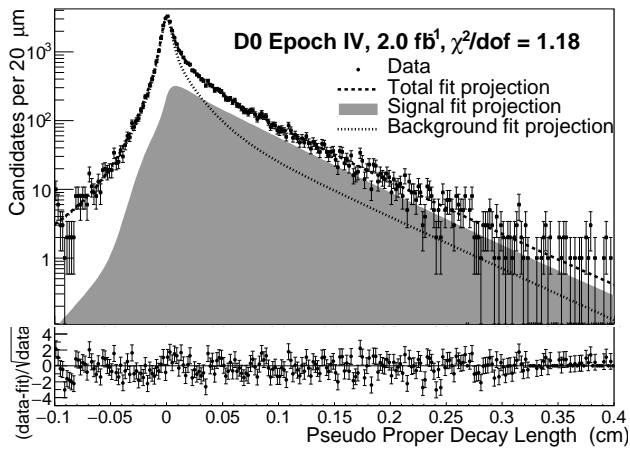}
\caption{Top: PPDL distribution for $D_s^-\mu^+$ candidates in the signal sample
for one of the five data periods.
The projections of the lifetime fitting model,
the background function, 
and the signal function are superimposed. Bottom: fit residuals demonstrating the agreement between the data and the fit model.}
\label{fig2}
\end{figure}

The combinatorial background PDF, $\mathcal{F}_{\text{comb}}$, is chosen empirically 
to provide a good fit to the combinatorial background PPDL distribution.
It is defined as the sum of the double-Gaussian resolution function 
 and two exponential decay functions for both the positive and negative PPDL regions.
The shorter-lived exponential decays are fixed to have the same slope for positive and negative regions,
while different slopes are allowed for the longer-lived exponential decays.
Figure ~\ref{fig2} shows the PPDL distribution for one of the five data periods for the signal sample, 
along with the comparison with the fit model. The corresponding $\chi^2$ per degree of freedom for 
each data-taking period are: $1.58, 1.21, 1.29, 1.18,$ and $1.14$.

The corresponding $B^0$ lifetime measurement uses exactly the same procedure for events in the $D^-$ mass peak, 
including a calculation of dedicated $K$ factors and background contributions from semileptonic decays. 

The lifetime fitting procedure is tested using MC pseudoexperiments, 
in which the generated $B_{(s)}^0$ lifetime is set to a range of different values, 
and the full fit performed on the simulated data.
Good agreement is found between the input and extracted lifetimes in all cases.
As an additional cross-check, the data are divided into pairs of subsamples, 
and the fit is performed separately for both samples. 
The divisions correspond to low and high $p_T(B_{(s)}^0)$, central and forward  pseudorapidity $|\eta(B_{(s)}^0)|$ regions, 
and $B_{(s)}^0$ versus $\bar{B}_{(s)}^0$ decays.
In all cases the measured lifetimes are consistent within uncertainties.

To evaluate systematic uncertainties on the measurements of $c\tau({B_s^0})$, 
$c\tau({B^0})$, and the ratio $\tau_{\rm{fs}}({B_s^0})/\tau({B^0})$, we consider 
the following possible sources: 
modeling of the decay length resolution, combinatorial background evaluation, 
$K$-factor determination, background contribution from charm semileptonic decays, signal fraction, and alignment of the detector. 
All other sources investigated are found to be negligible.
The effect of possible mismodeling of the decay length resolution is tested by 
repeating the lifetime fit with alternative resolution models, using a single 
Gaussian component. A systematic uncertainty is assigned based on the shift in the measured lifetime.
We repeat the fit using different combinatorial background samples 
using only the sideband data or only the wrong-sign sample.
The maximum deviation from the central lifetime measurement is assigned as a systematic uncertainty.
To determine the effect of uncertainties on the $K$ factors for the signal events, 
the fractions of the different components are varied within their uncertainties given in Table~\ref{table1}.
We also recalculate the $K$ factors using different MC decay models~\cite{evtgen}
leading to a harder $p_T$ distribution of the generated $B$ hadrons. 
The fraction of each component from semileptonic decays is
varied within its uncertainties, and the shift in the measured lifetime is used to assign a systematic uncertainty.
The signal fraction parameter, $f_{D_s\mu}$, is fixed for each mass fit performed.
We vary this parameter within its statistical and systematic uncertainty, obtained from fit variations to the background and signal
model of the mass PDFs, and assign the observed deviation as the uncertainty arising from this source.  
Finally, to assess the effect of possible detector misalignment, a single MC sample is 
passed through two different reconstruction algorithms, corresponding to the nominal detector alignment 
and an alternative model with tracking detector elements shifted spatially within their uncertainties.
The observed change in the lifetime is taken as systematic uncertainty due to alignment.

Table~\ref{TableSummary} lists the contributions to the systematic uncertainty from 
all sources considered. The most significant effect comes from the combinatorial background determination.
Correlations in the systematic uncertainties for the $B_s^0$ and $B^0$ meson lifetimes
are taken into account when evaluating the effect on the lifetime ratio, where the $K$ factor determination dominates.
               
\begin{table}[htb]
\begin{tabularx}{\columnwidth}{lcc>{\raggedleft\arraybackslash}X}
\hline\hline                
Uncertainty source            & $\Delta (c\tau_{B_s^0}) \mu\rm{m}$ & $\Delta (c\tau_{B^0}) \mu\rm{m}$ & ~~$\Delta R$~~\\[2pt]
\hline                
Resolution                    & $0.7$                 &  $2.1$              & $0.003$ \\                
Combinatorial background      & $5.0$                 &  $4.9$              & $0.001$ \\                
$K$ factor                    & $1.6$                 &  $1.3$              & $0.006$ \\                
Semileptonic components       & $2.6$                 &  $2.0$              & $0.001$ \\                
Signal fraction               & $1.0$                 &  $1.8$              & $0.002$ \\
Alignment of the detector     & $2.0$                 &  $2.0$              & $0.000$ \\
Total                         & $6.3$                 &  $6.4$              & $0.007$ \\
\hline\hline                
\end{tabularx}
\caption{Summary of systematic uncertainty contributions to the $B_s^0$ and $B^0$ lifetimes, 
and to the ratio $R \equiv \tau_{\rm{fs}}({B_s^0})/\tau({B^0})$.}
\label{TableSummary}                
\end{table}
      
The measured flavor-specific lifetime of the $B_s^0$ meson is 
$c\tau_{\rm{fs}}({B_s^0})  =  443.3 \pm 2.9   \thinspace{\rm(stat)} \pm 6.3   \thinspace{\rm(syst)}\thinspace\mu\rm{m}$,
which is consistent with the current world average of 
$439.2 \pm 9.3\thinspace\mu\rm{m}$~\cite{pdg,hfag}  and has a smaller total uncertainty of $6.9\thinspace\mu\rm{m}$.
The uncertainty in this measurement is dominated by systematic effects.
The $B^0$ lifetime in the semileptonic decay $B^0\rightarrow D^-\mu^+\nu X$
is measured to be 
$c\tau({B^0})  =  459.8 \pm 5.6   \thinspace{\rm(stat)} \pm 6.4  \thinspace{\rm(syst)}\thinspace\mu\rm{m}$,
consistent with the world average of $c\tau({B^0})= 455.4\pm 1.5\thinspace\mu\rm{m}$~\cite{pdg}.
Using both lifetimes obtained in the current analysis, their ratio is determined to be
$\tau_{\rm{fs}}({B_s^0})/\tau({B^0}) = 0.964 \pm 0.013 \thinspace{\rm(stat)} \pm 0.007 \thinspace{\rm(syst)}$. 
Both results are in reasonable agreement with theoretical
predictions from lattice QCD~\cite{theory1,theory2}, the flavor-specific lifetime has a better 
precision than the current world average~\cite{pdg,hfag}, and 
agrees reasonably well with the slightly more precise
recent measurement from the LHCb Collaboration~\cite{lhcb2}.

In summary, we measure the $B_s^0$ lifetime in the inclusive semileptonic channel
$B_s^0\rightarrow D_s^-\mu^+\nu X$ and obtain one of the most precise determinations 
of the flavor-specific $B_s^0$ lifetime.  Combining this result and that of Ref.~\cite{lhcb2} 
with global fits of  lifetime measurements in $B_s^0 \to J/\psi K^+  K^-$ decays~\cite{hfag} 
gives the most precise determination of the fundamental parameters $\Delta\Gamma_s$ and $\Gamma_s$
which are important for constraining $CP$ violation in the $B_s^0$ system.  Our precise 
measurement of the ratio of $B_s^0$ and $B^0$ lifetimes can be used to test and refine 
theoretical QCD predictions and offers a sensitive test of new physics~\cite{bobeth}. 

\input acknowledgement.tex
\end{document}

%% file: author_list.tex
\affiliation{LAFEX, Centro Brasileiro de Pesquisas F\'{i}sicas, Rio de Janeiro, Brazil}
\affiliation{Universidade do Estado do Rio de Janeiro, Rio de Janeiro, Brazil}
\affiliation{Universidade Federal do ABC, Santo Andr\'e, Brazil}
\affiliation{University of Science and Technology of China, Hefei, People's Republic of China}
\affiliation{Universidad de los Andes, Bogot\'a, Colombia}
\affiliation{Charles University, Faculty of Mathematics and Physics, Center for Particle Physics, Prague, Czech Republic}
\affiliation{Czech Technical University in Prague, Prague, Czech Republic}
\affiliation{Institute of Physics, Academy of Sciences of the Czech Republic, Prague, Czech Republic}
\affiliation{Universidad San Francisco de Quito, Quito, Ecuador}
\affiliation{LPC, Universit\'e Blaise Pascal, CNRS/IN2P3, Clermont, France}
\affiliation{LPSC, Universit\'e Joseph Fourier Grenoble 1, CNRS/IN2P3, Institut National Polytechnique de Grenoble, Grenoble, France}
\affiliation{CPPM, Aix-Marseille Universit\'e, CNRS/IN2P3, Marseille, France}
\affiliation{LAL, Universit\'e Paris-Sud, CNRS/IN2P3, Orsay, France}
\affiliation{LPNHE, Universit\'es Paris VI and VII, CNRS/IN2P3, Paris, France}
\affiliation{CEA, Irfu, SPP, Saclay, France}
\affiliation{IPHC, Universit\'e de Strasbourg, CNRS/IN2P3, Strasbourg, France}
\affiliation{IPNL, Universit\'e Lyon 1, CNRS/IN2P3, Villeurbanne, France and Universit\'e de Lyon, Lyon, France}
\affiliation{III. Physikalisches Institut A, RWTH Aachen University, Aachen, Germany}
\affiliation{Physikalisches Institut, Universit\"at Freiburg, Freiburg, Germany}
\affiliation{II. Physikalisches Institut, Georg-August-Universit\"at G\"ottingen, G\"ottingen, Germany}
\affiliation{Institut f\"ur Physik, Universit\"at Mainz, Mainz, Germany}
\affiliation{Ludwig-Maximilians-Universit\"at M\"unchen, M\"unchen, Germany}
\affiliation{Panjab University, Chandigarh, India}
\affiliation{Delhi University, Delhi, India}
\affiliation{Tata Institute of Fundamental Research, Mumbai, India}
\affiliation{University College Dublin, Dublin, Ireland}
\affiliation{Korea Detector Laboratory, Korea University, Seoul, Korea}
\affiliation{CINVESTAV, Mexico City, Mexico}
\affiliation{Nikhef, Science Park, Amsterdam, the Netherlands}
\affiliation{Radboud University Nijmegen, Nijmegen, the Netherlands}
\affiliation{Joint Institute for Nuclear Research, Dubna, Russia}
\affiliation{Institute for Theoretical and Experimental Physics, Moscow, Russia}
\affiliation{Moscow State University, Moscow, Russia}
\affiliation{Institute for High Energy Physics, Protvino, Russia}
\affiliation{Petersburg Nuclear Physics Institute, St. Petersburg, Russia}
\affiliation{Instituci\'{o} Catalana de Recerca i Estudis Avan\c{c}ats (ICREA) and Institut de F\'{i}sica d'Altes Energies (IFAE), Barcelona, Spain}
\affiliation{Uppsala University, Uppsala, Sweden}
\affiliation{Taras Shevchenko National University of Kyiv, Kiev, Ukraine}
\affiliation{Lancaster University, Lancaster LA1 4YB, United Kingdom}
\affiliation{Imperial College London, London SW7 2AZ, United Kingdom}
\affiliation{The University of Manchester, Manchester M13 9PL, United Kingdom}
\affiliation{University of Arizona, Tucson, Arizona 85721, USA}
\affiliation{University of California Riverside, Riverside, California 92521, USA}
\affiliation{Florida State University, Tallahassee, Florida 32306, USA}
\affiliation{Fermi National Accelerator Laboratory, Batavia, Illinois 60510, USA}
\affiliation{University of Illinois at Chicago, Chicago, Illinois 60607, USA}
\affiliation{Northern Illinois University, DeKalb, Illinois 60115, USA}
\affiliation{Northwestern University, Evanston, Illinois 60208, USA}
\affiliation{Indiana University, Bloomington, Indiana 47405, USA}
\affiliation{Purdue University Calumet, Hammond, Indiana 46323, USA}
\affiliation{University of Notre Dame, Notre Dame, Indiana 46556, USA}
\affiliation{Iowa State University, Ames, Iowa 50011, USA}
\affiliation{University of Kansas, Lawrence, Kansas 66045, USA}
\affiliation{Louisiana Tech University, Ruston, Louisiana 71272, USA}
\affiliation{Northeastern University, Boston, Massachusetts 02115, USA}
\affiliation{University of Michigan, Ann Arbor, Michigan 48109, USA}
\affiliation{Michigan State University, East Lansing, Michigan 48824, USA}
\affiliation{University of Mississippi, University, Mississippi 38677, USA}
\affiliation{University of Nebraska, Lincoln, Nebraska 68588, USA}
\affiliation{Rutgers University, Piscataway, New Jersey 08855, USA}
\affiliation{Princeton University, Princeton, New Jersey 08544, USA}
\affiliation{State University of New York, Buffalo, New York 14260, USA}
\affiliation{University of Rochester, Rochester, New York 14627, USA}
\affiliation{State University of New York, Stony Brook, New York 11794, USA}
\affiliation{Brookhaven National Laboratory, Upton, New York 11973, USA}
\affiliation{Langston University, Langston, Oklahoma 73050, USA}
\affiliation{University of Oklahoma, Norman, Oklahoma 73019, USA}
\affiliation{Oklahoma State University, Stillwater, Oklahoma 74078, USA}
\affiliation{Brown University, Providence, Rhode Island 02912, USA}
\affiliation{University of Texas, Arlington, Texas 76019, USA}
\affiliation{Southern Methodist University, Dallas, Texas 75275, USA}
\affiliation{Rice University, Houston, Texas 77005, USA}
\affiliation{University of Virginia, Charlottesville, Virginia 22904, USA}
\affiliation{University of Washington, Seattle, Washington 98195, USA}
\author{V.M.~Abazov} \affiliation{Joint Institute for Nuclear Research, Dubna, Russia}
\author{B.~Abbott} \affiliation{University of Oklahoma, Norman, Oklahoma 73019, USA}
\author{B.S.~Acharya} \affiliation{Tata Institute of Fundamental Research, Mumbai, India}
\author{M.~Adams} \affiliation{University of Illinois at Chicago, Chicago, Illinois 60607, USA}
\author{T.~Adams} \affiliation{Florida State University, Tallahassee, Florida 32306, USA}
\author{J.P.~Agnew} \affiliation{The University of Manchester, Manchester M13 9PL, United Kingdom}
\author{G.D.~Alexeev} \affiliation{Joint Institute for Nuclear Research, Dubna, Russia}
\author{G.~Alkhazov} \affiliation{Petersburg Nuclear Physics Institute, St. Petersburg, Russia}
\author{A.~Alton$^{a}$} \affiliation{University of Michigan, Ann Arbor, Michigan 48109, USA}
\author{A.~Askew} \affiliation{Florida State University, Tallahassee, Florida 32306, USA}
% ...
%
%
\collaboration{The D0 Collaboration\footnote{with visitors from
$^{a}$Augustana College, Sioux Falls, SD, USA,
$^{b}$The University of Liverpool, Liverpool, UK,
$^{c}$DESY, Hamburg, Germany,
$^{d}$Universidad Michoacana de San Nicolas de Hidalgo, Morelia, Mexico
$^{e}$SLAC, Menlo Park, CA, USA,
$^{f}$University College London, London, UK,
$^{g}$Centro de Investigacion en Computacion - IPN, Mexico City, Mexico,
$^{h}$Universidade Estadual Paulista, S\~ao Paulo, Brazil,
$^{i}$Karlsruher Institut f\"ur Technologie (KIT) - Steinbuch Centre for Computing (SCC),
D-76128 Karlsruhe, Germany,
$^{j}$Office of Science, U.S. Department of Energy, Washington, D.C. 20585, USA,
$^{k}$American Association for the Advancement of Science, Washington, D.C. 20005, USA,
$^{l}$Kiev Institute for Nuclear Research, Kiev, Ukraine,
$^{m}$University of Maryland, College Park, Maryland 20742, USA
and
$^{n}$European Orgnaization for Nuclear Research (CERN), Geneva, Switzerland
}} \noaffiliation
\vskip 0.25cm

%% file: acknowledgement.tex
%

We thank the staffs at Fermilab and collaborating institutions,
and acknowledge support from the
Department of Energy and National Science Foundation (United States of America);
Alternative Energies and Atomic Energy Commission and
National Center for Scientific Research/National Institute of Nuclear and Particle Physics  (France);
Ministry of Education and Science of the Russian Federation, 
National Research Center ``Kurchatov Institute" of the Russian Federation, and 
Russian Foundation for Basic Research  (Russia);
National Council for the Development of Science and Technology and
Carlos Chagas Filho Foundation for the Support of Research in the State of Rio de Janeiro (Brazil);
Department of Atomic Energy and Department of Science and Technology (India);
Administrative Department of Science, Technology and Innovation (Colombia);
National Council of Science and Technology (Mexico);
National Research Foundation of Korea (Korea);
Foundation for Fundamental Research on Matter (The Netherlands);
Science and Technology Facilities Council and The Royal Society (United Kingdom);
Ministry of Education, Youth and Sports (Czech Republic);
Bundesministerium f\"{u}r Bildung und Forschung (Federal Ministry of Education and Research) and 
Deutsche Forschungsgemeinschaft (German Research Foundation) (Germany);
Science Foundation Ireland (Ireland);
Swedish Research Council (Sweden);
China Academy of Sciences and National Natural Science Foundation of China (China);
and
Ministry of Education and Science of Ukraine (Ukraine).

%% file: d0-bs-lifetime.bbl
\begin{thebibliography}{99}
\bibitem{conjugate}  Charge conjugation is implied throughout this article.
\end{thebibliography}
